\documentclass[a4paper,11pt]{article}
\usepackage{jcappub} 
\usepackage{aas_macros}
\usepackage{lineno}
\usepackage{indentfirst}
\usepackage{enumerate}
\usepackage{booktabs}
\usepackage{multirow}
\usepackage[normalem]{ulem}
\usepackage{color,xcolor}
\def\h{\hat}
\def\t{\tilde}
\def\m{\mathrm}
\def\bm{\boldsymbol}
\useunder{\uline}{\ul}{}

\newcommand{\SectionRef}{Sec.~}
\newcommand{\SubSectionRef}{Subsec.~}
\newcommand{\TableRef}{Table.~}

\newcommand{\ukarcmin}{$\mu$K-arcmin }

\title{Forecasts on Anisotropic Cosmic Birefringence Constraints for CMB Experiment in the Northern Hemisphere}

\author[a,b]{Yiwei Zhong,}
\author[c,d,1]{Hongbo Cai,}
\author[e,1]{Si-Yu Li,}
\author[e]{Yang Liu,}
\author[f,g]{Mingzhe Li,}
\author[a,b,1]{Wenjuan Fang\note{Corresponding author.}}

\affiliation[a]{CAS Key Laboratory for Research in Galaxies and Cosmology, Department of Astronomy,\\
University of Science and Technology of China, Hefei, Anhui 230026, China}
\affiliation[b]{School of Astronomy and Space Sciences,\\
University of Science and Technology of China, Hefei, Anhui 230026, China}
\affiliation[c]{Department of Astronomy, School of Physics and Astronomy,\\
Shanghai Jiao Tong University, Shanghai, 200240, China}
\affiliation[d]{Key Laboratory for Particle Astrophysics and Cosmology (MOE)/Shanghai,\\
Key Laboratory for Particle Physics and Cosmology, Shanghai, China}
\affiliation[e]{Key Laboratory of Particle Astrophysics,\\
Institute of High Energy Physics, Chinese Academy of Sciences, Beĳing 100086, China}
\affiliation[f]{Interdisciplinary Center for Theoretical Study,\\
University of Science and Technology of China, Hefei, Anhui 230026, China}
\affiliation[g]{Peng Huanwu Center for Fundamental Theory,\\
University of Science and Technology of China, Hefei, Anhui 230026, China}

\emailAdd{zhongyiwei@mail.ustc.edu.cn}
\emailAdd{ketchup@sjtu.edu.cn}
\emailAdd{lisy@ihep.ac.cn}
\emailAdd{liuy92@ihep.ac.cn}
\emailAdd{limz@ustc.edu.cn}
\emailAdd{wjfang@ustc.edu.cn}

\abstract{
The study of cosmic birefringence through Cosmic Microwave Background (CMB) experiments is a key research area in cosmology and particle physics, providing a critical test for Lorentz and CPT symmetries. 
This paper focuses on an upcoming CMB experiment in the mid-latitude of the Northern Hemisphere, and investigates the potential to detect anisotropies in cosmic birefringence. 
Applying a quadratic estimator on simulated polarization data, we reconstruct the power spectrum of anisotropic cosmic birefringence successfully and estimate constraints on the amplitude of the spectrum, $A_{\mathrm{CB}}$, assuming scale invariance. 
The forecast is based on a wide-scan observation strategy during winter, yielding an effective sky coverage of approximately 23.6\%. 
We consider two noise scenarios corresponding to the short-term and long-term phases of the experiment. Our results show that with a small aperture telescope operating at 95/150GHz, the $2\sigma$ upper bound for $A_{\m{CB}}$ can reach 0.017 under the low noise scenario when adopting the method of merging multi-frequency data in map domain, and merging multi-frequency data in spectrum domain tightens the limit by about 10\%.
A large-aperture telescope with the same bands is found to be more effective, tightening the $2\sigma$ upper limit to 0.0062. 
}

\begin{document}
\maketitle
\flushbottom

\section{Introduction}
\label{sec:introduction}
The Cosmic Microwave Background (CMB) holds a crucial role in the field of physical cosmology as it provides direct observation of the earliest state of the Universe.
In recent years, detecting Cosmic birefringence has gradually become a hot topic in the field of CMB observation. 
Cosmic birefringence refers to the rotation of the linear polarization plane of CMB photons as they propagate through the universe.
It arises naturally in several physical theories beyond the standard model, and, in particular, can be induced by the Chern-Simons coupling between photons and the current of an external field, which reads:
\begin{equation}
    \mathcal{L}_{\text{CS}} \sim p_{\mu}A_{\nu} \tilde{F}^{\mu\nu}, 
\end{equation}
where $\tilde{F}^{\mu\nu}=\frac{1}{2} \epsilon^{\mu\nu \rho\sigma}F_{\rho\sigma}$ is the dual of the electromagnetic tensor.
This coupling term remains gauge invariant if $p_{\mu}$ is either a spacetime invariant or the derivative of a cosmological scalar field $\varphi$, which may be associated with a dark energy field \cite{2002PhRvD..65j3511L, 2003PhLB..573...20L}, a Ricci scalar field in baryo-/leptogenesis theory \cite{2004PhRvD..70d7302L,2004PhRvL..93t1301D}, or an axion-like field originating from string theory \cite{2010PhRvD..81l3530A}, topological defect theories \cite{2021JCAP...04..007T,2022JCAP...10..043K,2022JCAP...10..090J,2023PhLB..84337990G,2024JCAP...05..066F}, and other Lorentz-violating theories \cite{2017MPLA...3230002L}. 
Notably, if $p_{\mu}$ follows a non-zero background evolution, this Chern-Simons coupling term will break CPT symmetry. 
Therefore, measuring the cosmic birefringence effect would provide valuable hints in the search for new physics beyond the standard model and significantly aid in testing the CPT symmetry.

The Chern-Simons interaction mentioned above causes dispersion in left-handed and right-handed photons, leading to a rotation in the polarization direction. Consider  $p_{\mu} = \partial_{\mu}f(\varphi)$ for instance, the resulting rotation angle of the polarization direction for observed CMB photons can be determined through the following integral:
\begin{equation}
\alpha = \int_{\mathrm{O}}^{\mathrm{LSS}} \partial_{\mu}f(\varphi)dx^{\mu}(\lambda) =  f(\varphi_{\mathrm{LSS}}) - f(\varphi_{O})~, 
\end{equation}
where LSS denotes the Last Scattering Surface, and $\lambda$ represents the affine parameter along the path of light. 
The rotation angle can be split into two parts as $\alpha(\hat{\bm{n}}) = \bar{\alpha} + \delta \alpha (\hat{\bm{n}})$\cite{Li:2008tma}, where $\bar{\alpha}$ is the isotropic rotation angle and $\delta\alpha(\hat{\bm{n}})$ is the anisotropy, considered as a fluctuation with a zero mean $\langle\delta\alpha(\hat{\bm{n}})\rangle=0$.
Most of models predict both the isotropic and anisotropic cosmic birefringence, while some specific models such as massless scalar fields do not necessarily induce the isotropic one \cite{2011PhRvD..84d3504C}.


The isotropic polarization rotation angle induces odd-parity TB and EB power spectra \cite{Lue:1998mq, Feng:2004mq,Feng:2006dp}, which should vanish in standard theory without Chern-Simons modification. 
This phenomenon can be exploited for measurement purposes; however, it is fully degenerate with the universal miscalibration angle of the detectors' polarization orientation.
To break the degeneracy, a method was proposed in Refs.\cite{minami_2019,2020PTEP.2020j3E02M,2020PTEP.2020f3E01M}, which relies on the polarized Galactic foreground. 
Applying the method to nearly full-sky Planck data allows a signal of $\bar{\alpha}=0.35\pm  0.14^\circ$ at the 68\% confidence level \cite{2020PhRvL.125v1301M}. 
The precision on $\bar{\alpha}$ measurements has been improved with subsequent works \cite{2022PhRvL.128i1302D,2022A&A...662A..10E} and the latest one \cite{2022PhRvD.106f3503E} is $\bar{\alpha}={0.342^\circ}_{-0.091°}^{+0.094^{\circ}}$, implying a tantalizing hint of nonzero isotropic cosmic birefringence. 
Nevertheless, these results are sensitive to the model for the $EB$ power spectrum of the foregrounds, and a reliable one is still lacking at this time.

As to the anisotropic birefringence, it is usually described by the angular power spectrum of the polarization rotation angle, originating from the fluctuation of the external field. 
In \cite{Li:2008tma, Li:2013vga,Lee:2013mqa}, the authors adopted a non-perturbative expansion approach to obtain the relationship between the original and rotated CMB power spectra. where the fluctuation of external field still satisfies statistical isotropy. 
If we consider a specific realization of the polarization rotation pattern across the sky, it will break the statistical isotropy of the CMB, thus coupling the off-diagonal ($\ell \neq \ell'$) modes.
We can reconstruct the anisotropic rotation angle using the quadratic estimator technique \cite{2009PhRvD..80b3510G,2009PhRvL.102k1302K,2009PhRvD..79l3009Y}, in a manner similar to CMB lensing reconstruction \cite{2002ApJ...574..566H,2003PhRvD..67h3002O}. 
Many CMB experiments have carried out measurements on anisotropic cosmic birefringence, but no evidence has been found so far \cite{2012PhRvD..86j3529G,2015PhRvD..92l3509A,Liu:2016dcg,2017JCAP...12..046C,2017PhRvD..96j2003B,2020JCAP...11..066G,Namikawa_2020,Bianchini2020SPT,2022JCAP...09..075B,Bicep_line_of_sight}. 
The best constraint on the amplitude of a scale-invariant power spectrum at the 95\% confidence level is $A_{\text{CB}}\leqslant 0.044$ \cite{Bicep_line_of_sight} (See Eq.~\eqref{eq:CaaL} for the definition of $A_\mathrm{CB}$). Future CMB experiments are expected to improve this limit by orders of magnitude \cite{2019PhRvD.100b3507P}.
In addition to the power spectra, studies have also been conducted on the odd-parity CMB polarization power spectrum arisen from polarization rotation\cite{Kamionkowski:2010rb,Zhai:2020vob}.

Many experiments, especially the ground-based experiments as they can be easily scaled up with more detectors, have contributed to detecting anisotropic cosmic birefringence, such as POLARBEAR \cite{2015PhRvD..92l3509A}, ACTPol \cite{Namikawa_2020}, BICEP/Keck Array \cite{2017PhRvD..96j2003B,Bicep_line_of_sight}, SPTpol \cite{Bianchini2020SPT}, {Planck \cite{Zagatti:2024jxm}}, as well as the upcoming Simons Observatory \cite{2019JCAP...02..056A}, all of which are located at sites in Chile and Antarctica in the Southern Hemisphere.
In this paper, we will focus on the scientific potential of future high-precision Nortehern Hemisphere CMB experiments in probing the anisotropic cosmic birefringence. 
We based our forecast on AliCPT\cite{Li:2018rwc} as a representative, located at mid-latitudes in the Northern Hemisphere, with its first telescope AliCPT-1 being a small-aperture, dual-frequency instrument operating at 95 and 150 GHz.
In addition to the small-aperture telescope(SAT), we also considered a potential large aperture telescope(LAT) with the same frequency bands, which is expected to provide a substantial complement. 
We simulated data for the telescopes under various noise levels, then applied a quadratic estimator to construct the anisotropic polarization rotation maps, and estimated the amplitude of the rotation power spectrum.
When integrating estimators from the two bands, we adopted and compared two methods of joint analysis with multi-frequency data: map domain and spectrum domain combinations.

This paper is organized as follows: In \SectionRef\ref{sec:simulation}, we 
describe our simulation used for the anisotropic cosmic birefringence measurement forecast. In \SectionRef\ref{sec:methodology}, we introduce the methodology of rotation map and power spectrum estimation with single and multifrequency data. We present our forecast result in \SectionRef\ref{sec:result} and conclude in \SectionRef\ref{sec:conclusions}.

\textbf{Notation Remark}: The CMB fields with multipole moments $X_{\ell m},X\in \{E,B\}$ in this work involve both lensing and rotation effects.
Throughout this paper, we denote
rotation-induced quantities with prime, $X'_{\ell m}$ and lensing-induced
quantities with a tilde $\tilde{X}_{\ell m}$.

\section{Simulation}
\label{sec:simulation}

Linear polarization of CMB are charaterized by Stoke 
parameters $Q(\hat{\bm{n}})$ and $U(\hat{\bm{n}})$, where $\hat{\bm{n}}$ 
represents a unit vector pointing to the sky. $Q(\hat{\bm{n}})$ and $U(\hat{\bm{n}})$ jointly form a spin 2 or -2 vector, which can be decomposed into rotation-invariant E-mode and B-mode polarization patterns with spin-weighted spherical harmonics ${}_{\pm 2}Y^{*}_{\ell m}$ as
\begin{equation}
  \label{eq:EB lm}
  E_{\ell m} \pm iB_{\ell m} = -\int d^2\bm{\hat{n}}~{}_{\pm 2}Y^{*}_{\ell m} (Q \pm i U)(\bm{\hat{n}})~.
\end{equation}
With the cosmic birefringence, the direction of linear polarization of the primary CMB field is rotated according to
\begin{equation}
    (Q'\pm \mathrm{i} U')(\hat{\boldsymbol{n}})=(Q\pm \mathrm{i} U)(\hat{\boldsymbol{n}}) e^{\pm 2\mathrm{i} \alpha(\hat{\boldsymbol{n}})}~. \label{eq:QUrotate}
\end{equation}
In this work, we focus on anisotropic cosmic birefringence only, such that the isotropic part of rotation angle $\bar{\alpha} =0$ and we just use $\alpha$ to denote $\delta \alpha$ from now on.

The cosmic birefringence reconstruction pipeline relies on ensembles of simulations to estimate the mean field correction, correction in normalization, biases in raw spectrum of cosmic birefringence, and the uncertainties in the measured spectrum (See \SectionRef\ref{sec:methodology} for details). Our simulated sky maps include lensed (or lensed and rotated) CMB, foregrounds, and instrumental noise. Each realization includes observations at the 95GHz and 150GHz bands. The maps are produced with the \texttt{Healpix} \cite{healpix}\footnote{\href{http://healpix.sourceforge.net}{http://healpix.sourceforge.net}} pixelization scheme, adopting $N_{\mathrm{side}}=1024$.

The lensed CMB maps are produced using the \texttt{Lenspyx} \cite{lenspyx}\footnote{\href{https://github.com/carronj/lenspyx}{https://github.com/carronj/lenspyx}} package, with the 
best-fit Planck 2018 parameters \cite{planck2018param} as the input cosmology.
Simulations with rotation are also required for the purpose of pipeline validation, normalization factor calibration and bias terms evaluation. We mainly focus on the anisotropic rotation angle $\alpha$ described by a scale-invariant power spectrum
\begin{equation}
    \frac{L(L+1)}{2\pi}C_L^{\alpha\alpha}=A_{\mathrm{CB}}\times10^{-4}\text{ [rad}^2], \label{eq:CaaL}
\end{equation}
with an amplitude parameter $A_{\mathrm{CB}}$\footnote{Note that we follow the convention of $A_{\mathrm{CB}}$ as \cite{Namikawa_2020, Bianchini2020SPT}. It could be $10^4$ times of that defined in some other literature, where they adopted definition $\frac{L(L+1)}{2\pi} C^{\alpha \alpha}_L = A_{\mathrm{CB}}\ [\text{rad}^2]$}.
The rotation angle maps $\alpha(\hat{\boldsymbol{n}})$ are Gaussian realizations of a given power spectrum $C^{\alpha \alpha}_L$ and then the lensed CMB maps are rotated according to Eq.~\eqref{eq:QUrotate}. After that, the lensed and rotated CMB maps are transformed to obtain the spherical-harmonic space coefficients $X_{\ell m} (X \in \{E, B\})$, to which the foregrounds are added. Then the $X_{\ell m}$ coefficients are multiplied by the instrumental beam function and pixel window function. And finally the instrumental noises are added.

The foregrounds are Gaussian realizations with the underlying power spectrum obtained by fitting the \textit{d10s5} model of Python Sky Model \cite{pysm}, including thermal dust and synchrotron emission. The thermal dust \textit{d10} is modeled as a single component modified black body based on the Planck 2018 GNILC maps \cite{Planck_GNILC}, using spatially varying temperature $T_d$ and spectral index $\beta _d$. The synchrotron emission \textit{s5} uses WMAP 9-year 23 GHz Q/U map \cite{WMAP_9yr} as template in polarization, and is modeled as power law with a spatially varying spectral index $\beta _s$. The index $\beta _s$ is derived using a combination of the Haslam 408 MHz data and WMAP 23 GHz 7-year data \cite{WMAP_7yr_23GHz} and rescaled based on the S-PASS data \cite{S_PASS}.

We choose a certain elevation angle suitable for ground-based telescopes and, adopting a method similar to that used in \cite{Zhang:2023qxu}, generate the hitmap and corresponding white noise variance distribution based on site location and observing duration. The instrumental noises are Gaussian realizations generated from white noise variance maps. We trimmed portions with low signal-to-noise ratio near the edges of the survey footprint, then multiplied the Planck's foreground mask, and finally achieved a sky coverage of 23.6\%. To assess and compare the scientific capabilities of the experiment at different stages, we selected two different observation depths: high noise and low noise scenarios, with median values of map depths at 95 and 150 GHz being 5.6~\ukarcmin, 8.4~\ukarcmin and 4.2~\ukarcmin, 6.3~\ukarcmin  respectively. The high noise scenario corresponds roughly to 50 module-years of data accumulation for small-aperture AliCPT-1, while the low noise scenario is approximately 90 module-years. 
{Here, one module-year refers to the noise depth achieved by operating a single AliCPT-1 detector module (each module contains 852 detectors per frequency band) for one full observing season (from October to March, totaling approximately 2,000 hours).}
For a possible large aperture telescope LAT at the same site, we assume it adopts the same scanning strategy as AliCPT-1. Similarly, we considered two scenarios: high noise with a median noise level of 8.5~\ukarcmin and 12.7~\ukarcmin at 95/150GHz, corresponding to about 20 module-years of data accumulation, and low noise with 6.0~\ukarcmin and 9.0~\ukarcmin, corresponding to about 40 module-years. The experimental configurations are summarized in Table~\ref{tab:1}.
\begin{table}[]
    \centering
    \begin{tabular}{|c|c|c|c|c|}
\hline
Property & \multicolumn{2}{c|}{SAT} & \multicolumn{2}{c|}{LAT} \\
\hline
Aperture & \multicolumn{2}{c|}{72cm} & \multicolumn{2}{c|}{6m} \\
\hline
 Frequency & 95~GHz & 150~GHz & 95~GHz & 150~GHz \\
\hline
Beamsize & 19~arcmin & 11~arcmin & 2.3~arcmin & 1.4~arcmin \\
\hline
High noise $\sigma_n^P$ & 5.6 \ukarcmin & 8.4 \ukarcmin & 8.5 \ukarcmin & 12.7 \ukarcmin \\
\hline
Low noise $\sigma_n^P$ & 4.2 \ukarcmin & 6.3 \ukarcmin & 6.0 \ukarcmin & 9.0 \ukarcmin \\
\hline
\end{tabular}
    \caption{Summary of experimental configurations considered in this work. The $\sigma^{\mathrm{P}}_{n}$ is the median value of noise level.} 
    \label{tab:1}
\end{table}

\section{Methodology}
\label{sec:methodology}
In this section, we describe the methodologies of rotation map and power spectrum reconstruction given both single 
 frequency and multi-frequency data. In \SubSectionRef\ref{subsec: Analysis of Singlefrequency Data} and \SubSectionRef\ref{subsec:PS Estimation}, we
revisit respectively anisotropic cosmic birefringence field and power spectrum reconstruction using single frequency map.
In \SubSectionRef\ref{subsec: Joint Analysis of Multifrequency Data}, we introduce two strategies of joint analysis with multifrequency data as preparation for the realistic scenario in our forecast work.

\subsection{Rotation Field Reconstruction with Single Frequency Data}
\label{subsec: Analysis of Singlefrequency Data}
We consider rotation map reconstruction in the presence of both
CMB lensing and anisotropic cosmic birefringence. In this scenario, the CMB polarization fields to leading order of lensing potential $\phi$ and rotation field $\alpha$ can be described as
\begin{eqnarray}
  \label{eq:rot lens E}
    \t{E}'_{\ell m}&=&E_{\ell m}+\delta \t{E}_{\ell m}+ \delta E'_{\ell m}, \\
    \label{eq:rot lens B}
   \t{B}'_{\ell m}&=&\delta \t{B}_{\ell m}+ \delta B'_{\ell m},
\end{eqnarray}
where $E_{\ell m}$ refers to the multipole of primary $E$-mode, $\delta \t{E}_{\ell m}$, $\delta \t{B}_{\ell m}$ denote the first order perturbations from CMB lensing, and $\delta E'_{\ell m}$, $\delta B'_{\ell m}$ denote those from the rotation field. As our focus, the perturbations to the leading order of $\alpha$ in E-mode and B-mode polarization multipoles are
given by
\begin{equation}
  \label{eq:1st E and B rotation}
  \begin{aligned}
  \delta E'_{\ell m} &= -2\sum_{LM}\sum_{\ell'm'} (-1)^{m}\alpha_{LM} \left(\begin{array}{ccc}
\ell & L & \ell' \\
-m & M & m'
\end{array}\right) {}_{2} F^{\alpha}_{\ell L \ell'} \beta_{\ell L \ell'} E_{\ell' m'},\\
  \delta B'_{\ell m} &=\ 2\sum_{LM}\sum_{\ell'm'} (-1)^{m}\alpha_{LM} \left(\begin{array}{ccc}
\ell & L & \ell' \\
-m & M & m'
\end{array}\right)
{}_{2} F^{\alpha}_{\ell L \ell'} \epsilon_{\ell L \ell'} E_{\ell' m'},
\end{aligned}
\end{equation}
where $\alpha_{LM}$ is the rotation spherical harmonics multipoles given by 
\begin{equation}
    \alpha(\hat{\boldsymbol{n}})=\sum_{LM} \alpha_{LM} Y_{LM}(\hat{\boldsymbol{n}}), 
\end{equation}
$\epsilon_{\ell L \ell'}$ and $\beta_{\ell L \ell'}$ are parity terms defined as
\begin{equation}
  \label{eq:parity}
  \begin{aligned}
  &\beta_{\ell L \ell'}\equiv\frac{1-(-1)^{\ell+L+\ell'}}{2 i}, \\
&\epsilon_{\ell L \ell'}\equiv\frac{1+(-1)^{\ell+L+\ell'}}{2},
\quad
\end{aligned}
\end{equation}
and the function ${}_{2} F^{\alpha}_{\ell L \ell'}$ is defined as
\begin{equation}
{}_{2} F^{\alpha}_{\ell L \ell'} = \sqrt{\frac{\left(2 \ell+1\right)(2L+1)\left(2 \ell'+1\right)}{4\pi}}\left(\begin{array}{ccc}
\ell & L & \ell' \\
2 & 0 & -2
\end{array}\right)
\end{equation}
with the parentheses representing the Wigner-3j symbols. To leading order
in $\alpha$ and $\phi$, the off-diagonal covariance between E-mode and B-mode polarization fields is contributed by both lensing and rotation effects as

\begin{equation}
  \label{eq:rotated-lensed average}
  \begin{aligned}
\langle \t{E}'_{\ell m}\t{B}'_{\ell' m'} \rangle_{\mathrm{CMB}} = &\sum_{LM} \begin{pmatrix}
  \ell & \ell' & L\\
  m & m' & M
\end{pmatrix} f^{\phi}_{\ell L \ell'} \phi^{*}_{LM} \\
+&\sum_{LM} \begin{pmatrix}
  \ell & \ell' & L\\
  m & m' & M
\end{pmatrix} f^{\alpha}_{\ell L \ell'} \alpha^{*}_{LM},
  \end{aligned}
\end{equation}
for $\ell \neq \ell'$ and $m \neq -m'$, where $\phi_{LM}$ is the lensing potential multipole, and $\langle
... \rangle_{\m{CMB}}$ is defined to be an
ensemble average over different realizations of primary CMB with
fixed realizations of both $\phi$ and $\alpha$; $f^{\phi}_{\ell L \ell'}$
and $f^{\alpha}_{\ell L \ell'}$ are the weighting functions for $\phi$ and $\alpha$, and the latter one is given by
\begin{equation}
    \label{eq:weight function}
    f^{\alpha}_{\ell L \ell'}=-2\epsilon_{\ell L \ell'}{}_{2} F^{\alpha}_{\ell L \ell'}C^{\mathrm{EE}}_{\ell}.
\end{equation}

The reconstruction of rotation map takes advantage of
averaging off-diagonal covariance induced by $\alpha$ using a
quadratic estimator approach which is a similar methodology to
the reconstruction of $\phi$. The unormalized quadratic estimator for $\alpha$,  $\bar{\alpha}_{LM}$,
 with E and B maps is given by
 \begin{equation}
   \label{eq:unormalized estimator}
\bar{\alpha}_{LM}= \sum_{\ell _{1} m_{1}} \sum_{\ell _{2}m_{2}} (-1)^{M} \begin{pmatrix} \ell_1  & \ell _{2}  & L \\ m_{1}  &  m_{2}  &  -M\end{pmatrix} f_{\ell_{1} \ell_{2} L}^\alpha \frac{\hat{E}^{d}_{\ell_{1 }m_{1}}}{\hat{C}^{E E}_{\ell_{1}}} \frac{\hat{B}^{d}_{\ell_{2} m_{2}}}{\hat{C}^{BB}_{\ell_{1}}}
\end{equation}
where $\h{X}^{d}_{\ell m}$ ($X\in \{E, B\}$) are the multipoles from data; the filters in the denominators are the total observed power spectra
$\hat{C}^{XX}_{\ell_{1}} = \tilde{C}^{\mathrm{fid},XX}_\ell
+ N^{XX}_{\ell}$ where $\tilde{C}^{\mathrm{fid},XX}_\ell$ is
the fiducial lensed power spectrum for X without rotation effect and $N^{XX}_{\ell}$ is the
de-beamed noise spectrum. Note that as demonstrated in \cite{2009PhRvD..79l3009Y, Cai:2022zad},
in the appearance of lensing effect, the reconstruction of
$\alpha$ is not biased at the leading order. However, on the estimation of rotation power spectrum, lensing effect does induce a subdominant bias which is accounted for in \SubSectionRef\ref{subsec:PS Estimation}

This simplified diagonal filtering in Eq.~\eqref{eq:unormalized estimator} is slightly sub-optimal since it ignores small amounts of mode mixing due to masking, cross-correlation in $C^{TE}_{\ell}$ and foreground power spectrum.
However, it is significantly faster than optimal filtering used in \cite{Bianchini2020SPT} , which requires the use of conjugate-gradient methods \cite{Smith2007conjugate}. The input $E$ and $B$ modes with multipoles $200\leqslant \ell \leqslant 1200$ are used for SAT and $200\leqslant \ell \leqslant 2048$ for LAT. The modes with multipoles below $\ell_{\min} = 200$ are removed to mitigate foreground contamination following \cite{Namikawa_2020}. The $\ell_{\max}$ is chosen to account for the beam size of telescopes.

On the basis of Eq.~\eqref{eq:unormalized estimator}, the unbiased map estimator is given by
\begin{equation}
    \label{eq:normailized estimator}
   \h{\alpha}_{LM}=A_{L}(\bar{\alpha}_{LM}-\langle\bar{\alpha}_{LM}\rangle),
\end{equation}
where $A_{L}$ is a normalization factor, and $\langle\bar{\alpha}_{LM}\rangle$ is the mean-field bias. The mean-field bias accounts for the nonzero contribution to the estimator from inhomogeneous noise, beam asymmetry effects, mode-coupling indudced by mask, etc, and is estimated by averaging over the reconstructed maps using CMB simulations without rotation effect. {Specifically, We use 100 simulations to calculate the mean-field bias; the first 50 simulations are used to estimate the mean-field bias of the first $\hat{\alpha}$ that enter the cosmic birefringence spectrum calculation, and second 50 simulations to estimate the mean-field bias of the second $\hat{\alpha}$.} The analytical form of $A_{L}$, $A_{L}^{\mathrm{ana}}$, can be calculated by
\begin{equation}
A _L^{\m{ana}}= \left\{ \frac{1}{2L+1} \sum_{\ell_{1} \ell_{2}}  \frac{(f^{\alpha}_{\ell_{1} \ell_{2} L})^2}{\hat{C}^{E E}_{\ell_{1}} \hat{C}^{B B}_{\ell_{2}}} \right\} ^{-1}.
\end{equation}
Accounting for effects such as sky masking and filtering, the analytic normalization is corrected by a multiplicative factor of $\frac{\langle C^{\alpha^{\mathrm{in}} \alpha^{\mathrm{in}}}_{L} \rangle }{\langle C^{\alpha^{\mathrm{in}} \hat{\alpha}}_{L} \rangle }$ where 
$\langle C^{\alpha^{\mathrm{in}} \alpha^{\mathrm{in}}}_{L} \rangle$ is the average auto spectrum of the input rotation field realizations denoted by $\alpha^{\mathrm{in}}$ and $\langle C^{\alpha^{\mathrm{in}} \hat{\alpha}}_{L} \rangle$ is the average cross spectrum between the input $\alpha^{\mathrm{in}}_{LM}$ and reconstructed multipole $\hat{\alpha}_{LM}$ (with analytic normalization factor $A^{\mathrm{ana}}_L$ here).
The correction to $A^{\mathrm{ana}}_L$ is found to be $\lesssim 5\%$ in our simulation.

\subsection{Power Spectrum Estimation}
\label{subsec:PS Estimation}
Equipped with the reconstructed $\hat{\alpha}$, the cosmic birefringence power spectrum can be estimated by subtracting different types of noise biases from the raw power spectrum  
\begin{equation}
    C^{\hat{\alpha} \hat{\alpha}}_{L} = \frac{1}{w_4} \frac{1}{2L+1} \sum ^{L}_{M=-L} \hat{\alpha}_{LM} \hat{\alpha} ^{*}_{LM}, \label{eq:raw_spec}
\end{equation}
where $w_4$ is the average value of the fourth power of our sky mask. Among the noise biases, the most significant one
is the Gaussian noise N0 which encapsulates the 
disconnected contribution of the CMB four-point function at the level
of total power spectrum. N0 can be estimated by MC simulation \cite{Story2015} as
\begin{equation}
\label{eq:MC N0}
N^{(0)}_L  = \langle C_L^{\hat{\alpha} \hat{\alpha}}\left[\h{E}^{S_1} \h{B}^{S_2}, \h{E}^{S_1} \h{B}^{S_2}\right] \left.+C_L^{\hat{\alpha} \hat{\alpha}}\left[\h{E}^{S_1} \h{B}^{S_2}, \h{E}^{S_2} \h{B}^{S_1}\right]\right\rangle_{\mathrm{S_1}, \mathrm{S_2}},
\end{equation}
where we use the notation $C^{\hat{\alpha} \hat{\alpha}}_L[UV, XY]\equiv C^{\hat{\alpha}[UV]\hat{\alpha}[XY]}_L$ to demonstrate the dependence on four input CMB fields. $\h{X}^{S_1}$ and $\h{X}^{S_2}$ 
($X\in \{E, B\}$) are two sets of independent lensed CMB 
MC simulations generated by fiducial model with noise 
realization added but without rotation effect, and 
$\langle ... \rangle_{\mathrm{S_1}, \mathrm{S_2}}$ refers to 
an average over the two sets of MC simulations. 

In practice, we apply a modified version of N0, RDN0, which is 
the realization-dependent N0 bias to estimate the Gaussian noise 
in the raw power spectrum, and it is calculated following the approach introduced in \cite{BiasHarden2013,BiashardenPolarization} as:
\begin{equation}
    \begin{aligned}
{}^{(\m{RD})}N^{(0)}_L & = \left\langle C_L^{\hat{\alpha} \hat{\alpha}}\left[\h{E}^{d} \h{B}^{S_1}, \h{E}^{d} \h{B}^{S_1}\right]\right.  +C_L^{\hat{\alpha} \hat{\alpha}}\left[\h{E}^{S_1} \h{B}^{\mathrm{d}}, \h{E}^{\mathrm{d}} \h{B}^{S_1}\right]
\\& +C_L^{\hat{\alpha} \hat{\alpha}}\left[\h{E}^{S_1} \h{B}^{d}, \h{E}^{S_1} \h{B}^{d}\right] +C_L^{\hat{\alpha} \hat{\alpha}}\left[\h{E}^{S_1} \h{B}^{d}, \h{E}^{S_1} \h{B}^{d}\right] \\
&-C_L^{\hat{\alpha} \hat{\alpha}}\left[\h{E}^{S_1} \h{B}^{S_2}, \h{E}^{S_1} \h{B}^{S_2}\right] \left.-C_L^{\hat{\alpha} \hat{\alpha}}\left[\h{E}^{S_1} \h{B}^{S_2}, \h{E}^{S_2} \h{B}^{S_1}\right]\right\rangle_{\mathrm{S_1}, \mathrm{S_2}},
\end{aligned}
\end{equation}
where $\h{X}^{d}$ ($X\in \{E, B\}$) refers to data maps. 
RDN0 preserves the
realization-dependent information from the observed data
which is lost in N0 MC simulation and is less sensitive to
errors in the covariance matrix such as the statistical
uncertainties from the limited size of simulation set used to
calculate the covariance matrix and errors from the mismatch
between the underlying fiducial model used to generate
the simulations and the true physical model. Furthermore,
RDN0 also suppresses the covariance between
band powers \cite{full_covariance,joint_analysis2003,without_noise_2021}. Note that the average of the ${}^{(\m{RD})}N^{(0)}_L$
over many CMB realizations reproduces the standard Gaussian
noise N0. {To estimate the RDN0 bias, 400 lensed CMB simulations are divided into 2 sets of equal size, labeled by $S_1$ and $S_2$ respectively.}
 
The other terms to be subtracted from the raw power spectrum are the N1 bias and the lensing bias 
which are subdominant compared with N0. The N1 bias arises from those 
connected terms in CMB four-point function which contain the 
first-order rotation power spectrum $C^{\alpha \alpha}_{L}$, and is
estimated with set C simulations following a similar procedure to that introduced in \cite{Story2015} for CMB lensing estimation as
\begin{equation}
\label{eq:N1}
    \begin{aligned}
        N^{(1)}_L = & \left< C_{L}^{\hat{\alpha }\hat{\alpha }}[\t{E}'^{S^{\alpha}_1} \t{B}'^{S^{\alpha}_2},\t{E}'^{S^{\alpha}_1} \t{B}'^{S^{\alpha}_2}]\right. \\
         & +C_{L}^{\hat{\alpha }\hat{\alpha }}[\t{E}'^{S^{\alpha}_1} \t{B}'^{S^{\alpha}_2},\t{E}'^{S^{\alpha}_2} \t{B}'^{S^{\alpha}_1}]\\
         &-C_L^{\hat{\alpha} \hat{\alpha}}\left[\t{E}'^{S_1} \t{B}'^{S_2}, \t{E}'^{S_1} \t{B}'^{S_2} \right] \\
            &-C_L^{\hat{\alpha} \hat{\alpha}}\left[\t{E}'^{S_1} \t{B}'^{S_2}, \t{E}'^{S_2} \t{B}'^{S_1}\right]\rangle_{S^{\alpha}_1, S^{\alpha}_2, S_1, S_2},
        \end{aligned}
\end{equation}
where $\t{X}'\in \{E, B\}$ are lensed-rotated CMB polarization 
maps, $(S_{1}^{\alpha},S_{2}^{\alpha})$ labels maps with 
same rotation realizations, and $(S_1,S_2)$ labels maps 
with different rotation realizations. {The N1 bias is calculated with 200 lensed and rotated noiseless simulations with different realizations of  primary CMB and lensing potential. The 200 simulations are then grouped into 100 pairs, and each pair has the same realization of $\alpha$.}

The lensing bias is caused by the lensing effect that
must be subtracted as shown in \cite{Nami2017}. It 
is estimated by segregating the lensing contribution using MC simulations as
\begin{equation}
\label{eq:Nlens}
    N^{\mathrm{Lens}}_L=\langle C_L^{\hat{\alpha} \hat{\alpha}} \left[\t{E}^{S} \t{B}^{S}, \t{E}^{S} \t{B}^{S}\right]   - N^{(0)}_L\rangle_{S},
\end{equation}
{The lensing bias is estimated with another set of 100 lensed simulations.}

\begin{figure}[!htb]
    \center
    \includegraphics[width=0.8\linewidth]{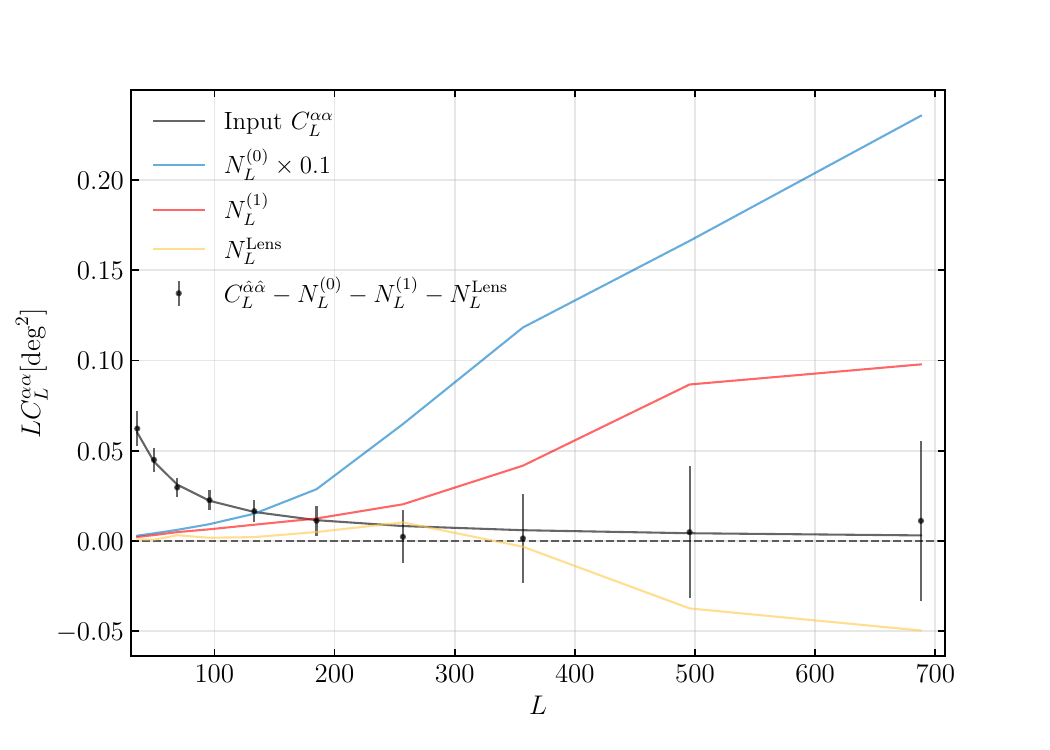}
    \caption{ Power spectrum reconstruction pipeline verification with small-aperture AliCPT-1 95 GHz channel in the high noise level scenario. The black line is the binned theoretical input cosmic birefringence spectrum \eqref{eq:CaaL} with $A_{\mathrm{CB}}=1$. The black dots represent the mean bandpowers from 400 {reconstructed rotation power spectrum}, after accounting for the bias terms $N^{(0)}  , N^{(1)}$ and $N^{\mathrm{Lens}}$ , and it reproduces the input spectrum within $1\sigma$. The error bars show the standard deviation of the {400 reconstructed power spectra}}
    \label{fig:fig1}
\end{figure}

Our final estimator for rotation power spectrum is 
\begin{equation}
\label{eq:ps estimator}
    \hat{C}^{\alpha\alpha}_L= C^{\hat{\alpha} \hat{\alpha}}_L - {}^{(\m{RD})}N^{(0)}_L - N^{(1)}_L - N^{\mathrm{Lens}}_L.
\end{equation}
To validate the power spectrum reconstruction pipeline with single frequency maps, we compare 
the averaged reconstructed rotation power spectrum with the 
input theoretical power spectrum as shown in Fig.~\ref{fig:fig1}. 
The averaged reconstructed rotation power spectrum is calculated by
$\langle C^{\hat{\alpha} \hat{\alpha}}_L\rangle - 
N^{(0)}_L - N^{(1)}_L - N^{\mathrm{Lens}}_L$ where 
$\langle C^{\hat{\alpha} \hat{\alpha}}_L\rangle$ is the mean bandpower of 400 raw power spectrum; note that $N^{(0)}_L$ is evaluated
using 
Eq.~\eqref{eq:MC N0} with a set of CMB realizations with rotation effect and thus is considered unbiased in estimating the total Gaussian noise; It is clear to see that the input theoretical rotation power spectrum can be recovered by our pipeline within $1 \sigma$.  



Note that when evaluating the covariance matrix, in principle, 
RDN0 should be subtracted for each 
simulation. However, it 
is impractical to perform such a computatinally expensive calculation since it involves averaging over hundreds of realizations of spectra obtained from different combinations of data and simulations for each RDN0. Instead, in the estimation of covariance matrix as we do in Sec.~\ref{sec:result} we adopt a faster semi-analytic approximation of RDN0 \cite{Planck2015lensing,act_dr6_lensing} given by:
\begin{equation}
    N^{(0),\mathrm{semi-analytic}}_L = \frac{A_L ^2 }{2L+1} \left[ \zeta_{L}(\hat{C}_\ell) - \zeta_{L}(C^{\mathrm{fid}}_\ell-\hat{C}_\ell) \right],
\end{equation}
where $\hat{C}_\ell$ is the observed CMB power spectrum and $C^{\mathrm{fid}}_{\ell}$ is the fiducial lensed power spectrum used in our simulations. The explicit form of $\zeta _L(C_\ell)$ is
\begin{equation}
    \zeta_{L}(C_\ell) = \sum _{\ell_1 \ell_2}\left[ (-1) ^{\ell_1+\ell_2+L} f^{\alpha}_{\ell_1 \ell_2 L} (F^{E}_{\ell_1} )^2  (F^{B}_{\ell_2})^2  C^{EE}_{\ell_1} C^{BB}_{\ell_2} + f^{\alpha}_{\ell_2 \ell_1 L }F^{E}_{\ell_1}F^{E}_{\ell_2} F^{B}_{\ell_1} F^{B}_{\ell_2} C^{EB}_{\ell_1} C^{EB}_{\ell_2}\right],
\end{equation}
where $F^X_\ell = (\tilde{C}^{\text{fid},XX}_\ell+ N^{XX}_\ell)^{-1}$. The semi-analytic approximation is only used in covariance matrix estimation since it is not accurate enough for power spectrum debiasing \cite{2020AA...641A...8P}. 

\subsection{Joint Analysis of Multifrequency Data}
\label{subsec: Joint Analysis of Multifrequency Data}
It is desirable to combine results from different frequencies to get more powerful constraints when considering the frequency-independent cosmic birefringence power spectrum. In this paper, we consider two approaches to combine results from different frequency bands.

The first one we adopted is the map-level inverse-variance coaddition strategy as described in \cite{act_map_combine}. The coadding of the maps is done in spherical-harmonic space, where maps from different frequency bands $D^{(f)}_{\ell m}$ are combined into a single CMB map $M_{\ell m}$: \begin{equation}
    M_{\ell m}=\sum_{f} w^{(f)}_\ell D^{(f)}_{\ell m} (B^{(f)}_{\ell}) ^{-1},
\end{equation}
where \begin{equation}
    w^{(f)}_{\ell}= \frac{(N^{(f)}_\ell) ^{-1} (B^{(f)}_\ell)^{2}}{\sum_{f} (N^{(f)}_\ell ) ^{-1}(B^{(f)}_\ell)^{2}}
\end{equation}
are normalized inverse variance weights, $N^{(f)}_{\ell}$ is the beam-deconvolved noise power spectrum and a deconvolution of the harmonic beam transfer function $B^{(f)}_{\ell}$ is performed for each frequency.

Another approach is to construct cosmic birefringence estimators with all possible combination of $E$- and $B$- modes, $\hat{\alpha}^{\hat{E}_i \hat{B}_j}$, where $i,j$ stand for 95 and 150 GHz channels respectively. With all combinations of $\hat{\alpha}^{\hat{E}_i \hat{B}_j}$, we then combine their auto- and cross-spectra and derive a joint constraint. With 4 individual estimators, we get a total of 10 spectra, where 4 of them are auto-spectra and 6 cross-spectra. The 10 possible auto- and cross-spectra are combined linearly with weights minimizing the variance of combined bandpowers:
\begin{equation}
    C^{\alpha\alpha}_{b} = \sum_{i} w_{b,i}C^{\alpha\alpha}_{b,i}
\end{equation}
where $i$ is the index for the 10 spectra and $b$ stands for the bins of the bandpowers. The weights are
\begin{equation}
    w_{b,i} =  \frac{ \sum _{j}\mathrm{Cov}^{-1}_{b,ij}}{\sum _{k j}\mathrm{Cov}^{-1}_{b,kj}},
\end{equation}
where $\mathrm{Cov}^{-1}_{b,ij}$ is the the covariance matrix of the bandpowers of bin $b$ from the 10 spectra.
\section{Results}
\label{sec:result}
\begin{figure}[htbp!]
        \centering
        \includegraphics[width=0.79\linewidth]{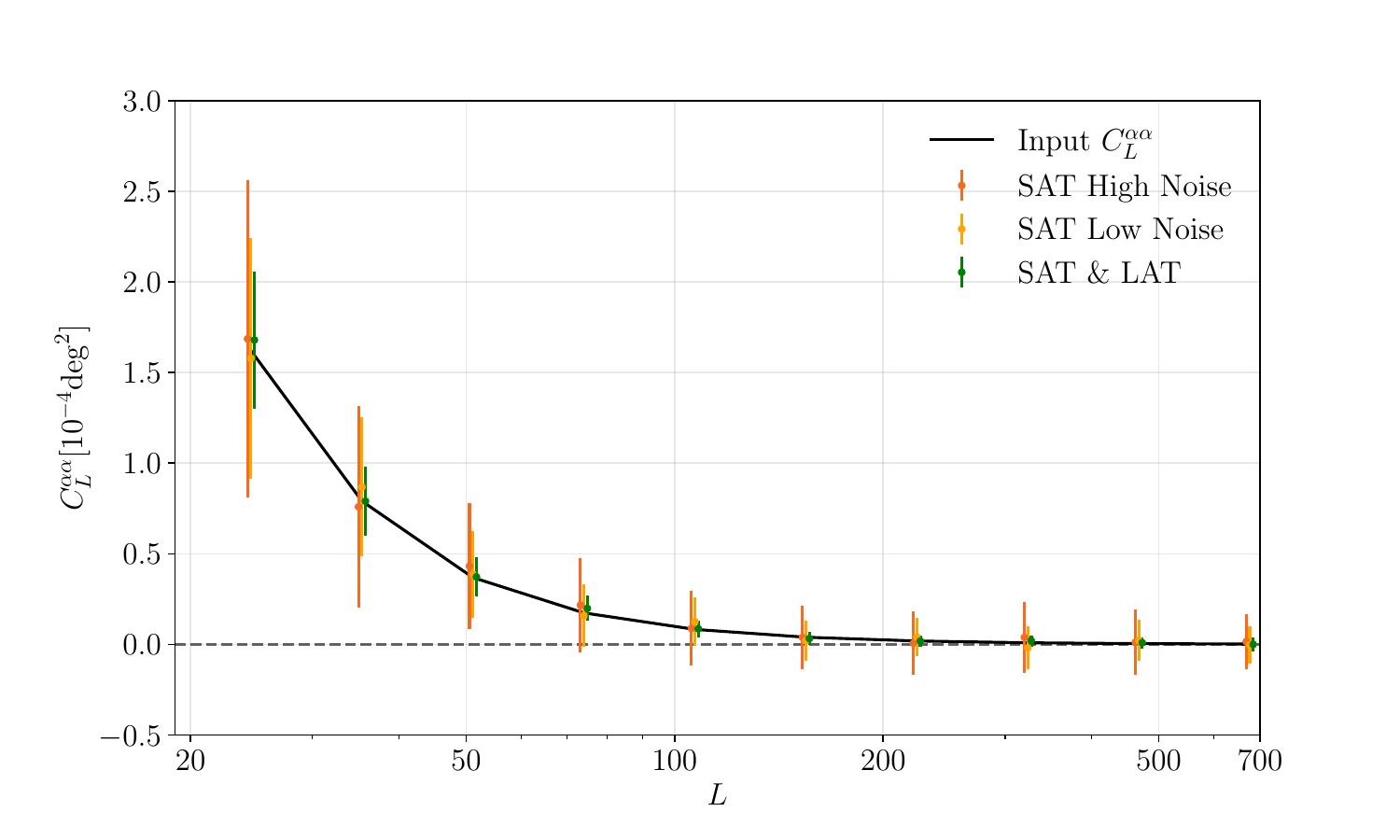}
        \includegraphics[width=0.8\linewidth]{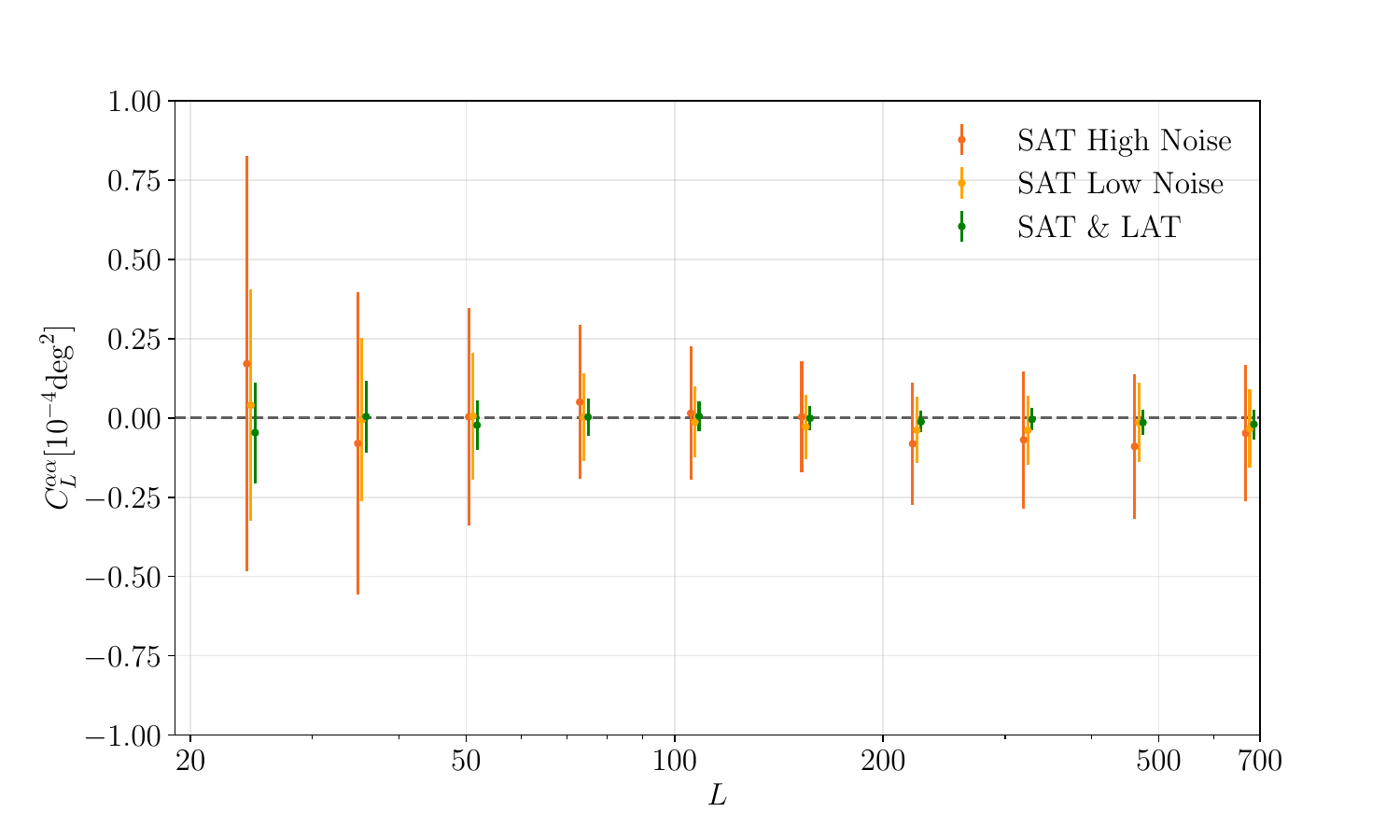}
        \caption{The upper panel shows the reconstructed cosmic birefringence power spectrum from simulations with a scale-invariant input power given by Eq.~\eqref{eq:CaaL} with $A_{\mathrm{CB}}=0.044$. The lower panel shows results from non-rotational simulations. All the results are obtained by combining two frequency bands at the map level. Here ``SAT \& LAT'' stands for results from combining the low noise level scenarios of both small-aperture AliCPT-1 and a potential large aperture telescope, at the map level. The error bars are estimated from the standard deviations among 400 {reconstructed power spectra}.}
        \label{fig:claa}
\end{figure}
In this section, we present the results of our analysis: 
the reconstructed rotation power spectrum using our 
multifrequency data pipeline given an input power spectrum, and the forecasts of constraints on the 
amplitude of scale-invariant rotation power spectrum with different 
experimental configurations.

We consider two scenarios for the input rotation 
power spectrum: 
one is a null test where the polarization rotation from cosmic birefringence is absent, the 
other is with a scale-invariant power spectrum whose 
amplitude is set to be the currently best $2\sigma$ 
upper bound $A_{\mathrm{CB}} = 0.044$ \cite{Bicep_line_of_sight}. In Figure~\ref{fig:claa},
we show the reconstructed rotation power spectra in these two scenarios using the map combine method with configurations of SAT (high noise), SAT (low noise), and the combination of SAT and LAT, with low noise for both of them.
The multipole range of spectra spans from 20 to 800 and 
is divided into 10 $\ell$-bins equally spaced in $\log{\ell}$.
The reconstructed power spectrum and corresponding error bars shown in the figure are the mean values and standard deviations from 400 simulations, respectively.
The average reconstructed power spectrum showed no significant bias compared with the
input well within $1\sigma$ for both scenarios. With the same input 
rotation power spectrum, we also apply the spectrum combine method
for the rotation power reconstruction, and find the differences of the standard deviations of these two methods are $\sim 10\%$ or less at each bin.

 
We then perform a comprehensive forecast for the sensitivity on 
the amplitude of the scale-invariant cosmic birefringence 
power spectrum $A_{\mathrm{CB}}$ by carving out the shape of 
the likelihood function. To adapt to the power spectrum in the 
largest bins which are not well described by Gaussian 
distributions, following 
\cite{Bianchini2020SPT,Namikawa_2020, Bicep_line_of_sight},
we employ a log likelihood proposed by Hamimeche and Lewis \cite{Hamimeche_Lewis_likelihood} given by
\begin{equation}
\label{eq:likelihood}
        \begin{aligned}-2\ln\mathcal{L}_{\alpha}(A_{\mathrm{CB}})=\sum_{bb^{\prime}}g(\hat{A}_{L_{b}})C_{L_{b}}^{f}\mathbb{C}_{L_{b}L_{b^{\prime}}}^{-1}C_{L_{b^{\prime}}}^{f}g(\hat{A}_{L_{b^{\prime}}}),\end{aligned}
\end{equation}
where
\begin{equation}
\hat{A}_L=\frac{\hat{C}_L^{\alpha\alpha, \mathrm{obs}}+N_L^0+N_L^{\text{lens}}}{ A _ {\text{CB}} ( C _ L ^ { \alpha \alpha , \mathrm{ref}}+N_L^1)+N_L^0+N_L^{\text{lens}}}~,
\end{equation}
is the amplitude of the recovered power spectrum relative to that of simulations including the rotation signal $C^{\alpha \alpha}_{L}$ at each bin $L_{b}$, and  
$g(x)=\mathrm{sign}(x-1) \sqrt{ 2(x-\ln x-1) }$ for $x\geq 0$ at 
a given bin $L_b$;
$\hat{C}_L^{\alpha\alpha, \mathrm{obs}}$ is the 
power spectrum reconstructed from observation; 
$C_L^{\alpha\alpha, \mathrm{ref}} = 2\pi / [L(L+1)]
\times 10^{-4}\ [\text{rad}^2]$ represents a reference scale-invariant power spectrum corresponding to $A_{\m{CB}}^{\m{ref}}=1$; the fiducial power spectrum $C_{L_{b}}^{f}$ and the covariance matrix $\mathbb{C}_{L_b L_b'}$ are both evaluated by unrotated simulations.  
Relevant noise terms are obtained in the way described in \SectionRef\ref{subsec:PS Estimation}. Note here $N^1_L$ is included as part of signal since it is proportional to $C^{\alpha\alpha}_{L}$.
For each simulation, the reconstructed power spectrum is estimated for 10 multipole bands in the range of $20\leqslant  L \leqslant 800$. 
In our forecast, for a given experimental configuration, we 
set $\hat{C}_L^{\alpha\alpha, \mathrm{obs}} =0$, and obtain 
the $2\sigma$ upper limit on $A_{\m{CB}}$ by adjusting 
the value of the likelihood to the corresponding 
$2\sigma$-level. This 
can be explained as how likely it is to observe  a null 
power spectrum on average with a given value of $A_{\m{CB}}$.


\begin{table}[htbp]
        \centering
        \begin{tabular}{llll}
        \toprule
        Instrument     & {Frequency / Co-added strategy}   & Noise  & $A_{\mathrm{CB}}$   \\ \midrule
        SAT       &    95 GHz            &    High            & 0.050  \\
                  &    95 GHz            &    Low            & 0.030  \\
                  &  150 GHz             &    High             & 0.050  \\
                  &  150 GHz             &    Low            & 0.028  \\
                  &   Map combine        &    High            & 0.029  \\
                  &   Map combine        &    Low           & 0.017  \\
                  &   Spectra combine    &    High           & 0.026  \\
                  &   Spectra combine    &    Low            & 0.016  \\ \midrule
     LAT &   Map combine        &    High            & 0.011  \\
                  &   Map combine        &    Low            & 0.0062 \\ \midrule
SAT \& LAT&  Map combine   &    Low    & 0.0055 \\
                \bottomrule
        \end{tabular}
        \caption{The forecasted $2\sigma$ upper bounds on $A_{\mathrm{CB}}$. Here SAT \& LAT means results from combining low noise level scenarios of both the small-aperture AliCPT-1 and a potential LAT. For details about the noise level, see \TableRef\ref{tab:1}.}
        \label{tab:Acb}
\end{table}

The forecasts for the $2\sigma$ upper bounds on $A_{\mathrm{CB}}$ for different experimental configurations are summarized in Table~\ref{tab:Acb}.
First, we calculate the sensitivity for each single band of small-aperture AliCPT-1, and find that the sensitivities at 95 and 150 GHz are comparable with each other.
Note that as mentioned in Sec.~\ref{sec:methodology}, the CMB multipoles below $\ell_{\m{min}}$ have been truncated to get the foreground removed. 
And at this point while the resolution at 95 GHz is slightly poorer, the lower noise level compensates for that.

For single band results of small-aperture telescope, we can also see an approximate-40\% better sensitivity at low noise levels compared to high noise levels, which is consistent with our expectation considering the nearly doubled data accumulation in the low noise scenario.
On top of that, we calculated the sensitivities when applying co-added strategies including both the map combine method and the spectrum combine method. 
We find that both the two co-added methods significantly improve the sensitivity compared with that of single-frequency bands by about 40\%-50\%. One can also see that spectrum combine method slightly wins out the map combine method, whereas the former takes about one to two orders of magnitude more computational resource compared with the the latter depending on the number of frequency bands involved.


Next, for the results of the LAT, we only present sensitivities forecasted for the map combine method considering the much less computational cost compared with the spectrum combine method. 
It is evident that the large-aperture telescope provides significant benefits due to its advantageous resolution.
 {Given} that in our forecast, the data amount under LAT's low noise scenario is roughly equivalent to that of small-aperture AliCPT-1's high noise scenario, the $2\sigma$ upper bound is remarkably tightened by about 80\%.
When combining the SAT and LAT, we obtain an even better 
sensitivity which gives the tightest $2\sigma$ upper limit of 0.0055 among all configurations we consider.
It should be noted that all the results we listed in the table are obtained after discarding the reconstructed power spectrum at $L\leqslant 20$, in order to avoid being affected by the isotropic polarization rotation or the uniform polarization miscalibration \cite{Namikawa_2020}. However, with further investigations of the systematic effects on the large-scale reconstructed rotation power spectrum (bin of $[2,20)$), we expect a stronger constraining power with this scale included, better by about a factor of a few according to our raw evaluation. 
 

\section{Conclusions}
\label{sec:conclusions}
Probing primordial gravitational waves using terrestrial CMB experiments is currently a prominent research focus in both cosmology and particle physics. 
In addition to this, detecting potential Chern-Simons interactions through the phenomenon of cosmic birefringence as a method to test Lorentz and CPT symmetries becomes an increasingly important objective for such experiments. 
In this paper, we focus on the upcoming CMB experiment in the Northern Hemisphere, and forecast for the prospective capabilities in detecting anisotropies of cosmic birefringence in the near future by taking the AliCPT project as a model.


We applied the minimum variance quadratic estimator on the mock data to reconstruct the power spectrum of the anisotropic birefringence, and thus we obtained the constraints on the spectrum's amplitude $A_{\mathrm{CB}}$ under a scale-invariant assumption.
Our forecast adopts the wide scan observation strategy during the winter months to achieve large sky coverage.
After masking the galactic plane and high noise region, the effective sky coverage is approximately 23.6\%. 
The large-scale multipoles ($\ell<200$) of the observational maps are removed, which ensures the forecasted results are affected by galactic foregrounds at minimum level. 
We considered two noise scenarios listed in \TableRef\ref{tab:1}, corresponding to the short-term and long-term phases of the experiment.
When merging bands or telescopes, two methods were applied: one directly merges the data in the map domain, and the other constructs estimators in each band separately and then combines them using a minimum variance approach.
The difference in results obtained with these two methods is less than 10\%.
The results show with the initial telescope of AliCPT, a dual-band small-aperture telescope AliCPT-1, the $2\sigma$ upper bounds on $A_{\mathrm{CB}}$ can reach 0.026 and 0.016 under high/low noise scenarios, respectively.

Note that for the forecast of $A_{\m{CB}}$ in our analysis, we do not adopt the multipoles with $L<20$ of the reconstructed rotation power spectrum in order to avoid the potential biases in this range like the isotropic miscalibration. With more comprehensive analysis of the systematic effects in this range in future work, we expect a prominent improvement by a factor of a few in the constraining power on anisotropic cosmic birefringence. 

Furthermore, our results indicate that a large-aperture telescope is more effective in constraining the power spectrum of anisotropic birefringence. 
Under high/low noise scenarios, the $2\sigma$ upper limits on $A_{\mathrm{CB}}$ could achieve 0.011 and 0.0062, respectively. 
The combination of two telescopes will further tighten the constraints.
\acknowledgments
We thank Jiazheng Dou, Chang Feng, Yilun Guan and Pengjie Zhang for helpful discussions. 
We thank Toshiya Namikawa for very useful suggestions. 
This work is supported by the National Key R\&D Program of China (2021YFC2203102, 2020YFC2201600, 2020YFC2201602, 2022YFF0503404, 2023YFA1607800, 2023YFA1607801), 
by the National Natural Science Foundation of China (12173036, 12075231, 11773024, 11621303), 
by the China Manned Space Project “Probing dark energy, modified gravity and cosmic structure formation by CSST multi-cosmological measurements” and grant No. CMS-CSST-2021-B01, CMS-CSST-2021-A02,
by the Fundamental Research Funds for Central Universities (WK3440000004 and WK3440000005), 
by Cyrus Chun Ying Tang Foundations, 
and by the 111 Project for “Observational and Theoretical Research on Dark Matter and Dark Energy” (B23042). 

\bibliographystyle{JHEP}
\bibliography{biblio.bib}

\end{document}